\documentclass[aps,pre,showkeys,preprint]{revtex4}
\usepackage{graphicx}
\usepackage{amsmath}
\usepackage{amsfonts}
\usepackage{amssymb}
\usepackage{bm}
\usepackage{enumitem}  
\usepackage{float}
\usepackage{xcolor}
\usepackage{color}

\begin{document}
\title{Electron Holes in a $\kappa$ Distribution Background with Singularities}
\author{Fernando Haas}
\affiliation{
Physics Institute, Federal University of Rio Grande do Sul, Av. Bento Gon\c{c}alves 9500, 91501-970 Porto Alegre, RS, Brazil\\
Email Address: fernando.haas@ufrgs.br}
\begin{abstract}
The pseudo-potential method is applied to derive diverse propagating electron hole structures, in a nonthermal or $\kappa$ particle distribution function background. The associated distribution function Ansatz reproduces the Schamel distribution of \cite{Schamel2015} in the Maxwellian ($\kappa \rightarrow \infty$) limit, providing a significant generalization of it for plasmas where superthermal electrons are ubiquitous, such as space plasmas.  The pseudo-potential and the nonlinear dispersion relation are evaluated. The role of the spectral index $\kappa$  on the nonlinear dispersion relation is investigated, in what concerns the wave amplitude for instance.  The energy-like first integral from Poisson's equation is applied to analyze the properties of diverse classes of solutions: with the absence of trapped electrons, with a non-analytic distribution of trapped electrons, or with a surplus of trapped electrons. Special attention is therefore paid to the non-orthodox case where the electrons distribution function exhibits strong singularities, being discontinuous or non-analytic.
\end{abstract}
\maketitle
\section{Introduction}

Although the usual treatment considers analytic distribution functions, more appropriate for 
quiescent plasmas, in noisy or turbulent plasmas such as fusion plasmas it can be expected to have 
some degree of singular distribution functions. Such systems require a nonlinear approach for which the 
derivation of coherent structures from nonlinear methods is a welcome task. 

Moreover in the last decades experiments have found the ubiquitous appearance of hole structures, 
for instance electron holes, solitary waves and double layers in space plasmas as in the free solar wind, at interplanetary
shocks \cite{r1, rr2, rr3, rr4} and collisionless laboratory plasmas \cite{r2}, 
as recently reviewed in \cite{hut}. In this context, Schamel has discussed stationary electrostatic waves propagating with 
a nonzero speed in a collisionless thermal plasma with singularities in the distribution function \cite{Schamel2015}. 
For this purpose the method employed was the pseudo-potential method, where at first the distribution function has 
a supposed form in terms of constants of motion, automatically solving the Vlasov equation thanks to Jean's theorem. 
Afterward the number density is evaluated as a function of the electric potential, up to a certain order so that the 
treatment is weakly nonlinear. Taking into account the Poisson equation it is possible to express the conditions 
for a localized solution in terms of a nonlinear dispersion relation. Solitary or periodic and cnoidal waves can be therefore 
described, with a focus on the impact of the singularities of the distribution function. Interestingly, the singular character 
of the distribution function with discontinuity at the separatrix or a non-analytic trapped electrons distribution do not 
transfer to the hole solutions, which are typically smooth. This is due to the fact that the singularities are somewhat 
washed when integrating in velocity space in order to obtain the charges number density, as apparent in Eq. (\ref{yy}) below.

However, frequently plasmas have not a Maxwellian equilibrium velocity distribution, having instead a power-law distribution above the thermal speed. This is a typical situation in both space and laboratory plasmas. Superthermal electrons are ubiquitous in the solar wind \cite{r5}, in Saturn's  magnetosphere \cite{r4}, in beam-plasma interactions \cite{r12} and intense laser-matter experiments \cite{r13}, besides numerical simulations \cite{r10}. These systems are better described by a $\kappa$ distribution (also called generalized Lorentzian distribution).

Our goal is to provide the generalization of \cite{Schamel2015}, which considering a singular $\kappa$ velocity distribution which reduces to a Maxwellian in the thermal limit $\kappa \rightarrow \infty$ equilibrium. Using the pseudo-potential method, the Sagdeev potential is derived with an emphasis on the impact of the singularities associated with trapping, and on the existence and behavior of diverse classes of hole solutions. As already remarked in \cite{Schamel2015}, a Maxwellian equilibrium tends to be more amenable to analytical calculations. We show here how to overcome the odds arising from a nonthermal equilibrium, thanks to the use of the generalized $\kappa$ plasma dispersion function (see Eq. (\ref{e13}) below). 

This work is organized as follows. In Section II the Schamel distribution function originally proposed for electrostatic waves in a thermal equilibrium is adapted to a $\kappa$ background. The appropriate rescaling to non-dimensional variables is applied. In Section III, the electrons number density is evaluated in the small amplitude limit, together with the corresponding pseudo-potential. The nonlinear dispersion relation compatible with localized structures is derived. In Section IV special classes of solutions are discussed, with an emphasis on the singular aspects of the trapping: solutions in the complete absence of trapping, with a non-analytic trapped electrons distribution, and with an excess of trapping. Section V is reserved to the conclusions.

\section{One-dimensional Superthermal Distribution}

Our starting point is the one-dimensional (1D) $\kappa$ distribution function for electrons,
\begin{equation}
\label{e1}
f(v) = \frac{n_0}{(\pi\kappa\theta^2)^{1/2}}\,\frac{\Gamma(\kappa)}{\Gamma(\kappa-1/2)}\,\left(1 + \frac{v^2}{\kappa\theta^2}\right)^{-\kappa} \, 
\end{equation}
where
\begin{equation}
\label{e2}
\theta^2 = \left(\frac{2\kappa-3}{\kappa}\right)\,\left(\frac{\kappa_B T}{m}\right) \,, \quad \kappa > 3/2 \,,
\end{equation}
as proposed in \cite{Summers, Hellberg2002, Podesta, Williams, Hellberg2009}. It appears from the three-dimensional (3D) $\kappa$ distribution after integration over two velocity components. Note the dependence on the inverse power of $\kappa$, while in the 3D version it is $\kappa + 1$. In Eqs. (\ref{e1}) and (\ref{e2}), $n_0$ is the equilibrium number density, $\Gamma$ is the gamma function, $\kappa$ is the spectral index, $\theta$ is the thermal speed, $m$ is the electron mass, $\kappa_B$ is the Boltzmann constant and $T$ is the temperature, as defined from the second moment of the distribution function, 
\begin{equation}
\label{e3}
\frac{1}{n_0}\int_{-\infty}^{\infty} f(v) v^2 dv = \frac{\kappa_B T}{m} \,.
\end{equation}

Our interest will be on propagating electrostatic structures, stationary in the wave frame.  
In this case, the stationary Vlasov equation, as is well known, is solved by a function of the constants of motion, namely
\begin{equation}
\label{e4}
\epsilon = \frac{m v^2}{2} - e\phi \,, \quad \sigma = {\rm sgn}(v) \,,
\end{equation}
where $-e$ is the electron charge and $\phi = \phi(x)$ is the scalar potential. The sign of the velocity $\sigma = \sigma(v)$ is a constant of motion for untrapped electrons. Without loss of generality, the separatrix separating passing and trapped electrons is set at $\epsilon = 0$. In addition, an homogeneous ionic background is also included, so that the Poisson equation reads 
\begin{equation}
\label{yy}
\frac{\partial^2\phi}{\partial x^2} = \frac{e}{\varepsilon_0}(n - n_0) \,, \quad n = \int_{-\infty}^{\infty} f(v) dv \,,
\end{equation}
where $\varepsilon_0$ is the vacuum permittivity.

From the $\kappa$ distribution (\ref{e1}), replacing $v \rightarrow \sqrt{2\epsilon/m}$ and after a few more adjustments, the shifted $\kappa$ distribution is then
\begin{eqnarray}
\label{e5}
f &=& \frac{n_0 \,(1 + k_0^2\Psi/2)}{(\pi\kappa\theta^2)^{1/2}}\,\frac{\Gamma(\kappa)}{\Gamma(\kappa-1/2)}\,\Bigl[H(\epsilon)\left(1 + \frac{1}{\kappa\theta^2}(\sigma \sqrt{\frac{2\epsilon}{m}} + v_0)^2\right)^{-\kappa} + \\
&+& \alpha H(-\epsilon)\left(1 + \frac{v_0^2}{\kappa\theta^2}\right)^{-\kappa}\left(1 + \gamma \sqrt{-\,\frac{\epsilon}{m\kappa\theta^2}} - \,\frac{\beta\,\epsilon}{m\kappa\theta^2}\right)\Bigr] \,,
\nonumber
\end{eqnarray}
where $v_0$ is the phase velocity in the electrons lab frame. 
It consists of two parts, the first one for untrapped electrons, the second for trapped electrons containing the parameters $\alpha, \beta$ and $\gamma$. In the limit of $\alpha = 1$, $\gamma = 0$ and small amplitudes, it would be the $\kappa$ version of the Schamel distribution \cite{Schamel2012} adapted to a Maxwellian equilibrium, apart from the normalization constant choice. At this point, $k_0$ and $\Psi$ are dimensionless variables, where $\Psi$ is proportional to the electrostatic potential amplitude and $k_0$ is related to the wavenumber of oscillatory solutions, whose role is to be better specified later. 

The distribution (\ref{e5}) exactly solves the stationary Vlasov equation and corresponds to the singular distribution shown in Eq. (2) of \cite{Schamel2015}, which is adapted to a Maxwellian background. The parameter $\alpha$ is a measure of the trapping strength, noting that $\alpha \neq 1$ imply a jump across the separatrix. Accordingly, $\alpha > 1$ is associated with overpopulated trapped electrons, while $\alpha < 1$ has the opposite meaning. For the trapped part, the more regular $\gamma = 0$ case imply an expansion in powers of $-\epsilon$ rather than in powers of $\sqrt{-\epsilon}$. Finally, $\beta$ represents a fine tuning of the inverse temperature of the trapped population. Our choice is justified to have a close resemblance with the singular equilibrium in a Maxwellian background of \cite{Schamel2015} but now with superthermal electrons. Certainly, higher singularities could be also included \cite{Schamel2020a, Schamel2020b} but here we keep to a bare minimum, for simplicity.
Notice that other $\kappa$ distributions used in studies of electron holes by means of the 
pseudo-potential method do not reduce to the choice of \cite{Schamel2015} and not only because they are regular, non-singular, but in view of an intrinsic different form. For instance, see Eqs. (1) and (2) of \cite{Abbasi2007}, where the electrons distribution function actually is not a function of the energy, or Eqs. (1) and (2) of \cite{Abbasi2008}, where it is non-propagating ($v_0 = 0$).

To proceed, it is convenient to rescale variables according to 
\begin{eqnarray}
\bar{x} &=& x/\lambda_D \,, \quad \bar{v} = v/v_T \,, \quad \bar{v}_0 = v_0/v_T \,, \quad \bar{\phi} = \frac{e\phi}{\kappa_B T} \,, \\
\bar{n} &=& n/n_0 \,, \quad \bar{f} = \frac{f}{n_0/v_T} \,, \quad \bar{\gamma} = \frac{\gamma}{\sqrt{2\kappa-3}} \,, \quad \bar{\beta} = \frac{\beta}{2\kappa-3} \,, 
\nonumber
\end{eqnarray}
where $v_T = \sqrt{\kappa_B T/m}$ and $\lambda_D = \sqrt{\epsilon_0\kappa_B T/n_0 e^2}$. With these choices and dropping bars from now on, we have 
\begin{eqnarray}
\label{e6}
f &=& A\,\Bigl(1 + \frac{k_0^2\Psi}{2}\Bigr)\,\Bigl[H(\epsilon)\left(1 + \frac{1}{2\kappa-3}(\sigma \sqrt{2\epsilon} + v_0)^2\right)^{-\kappa} + \\
&+& \alpha H(-\epsilon)\left(1 + \frac{v_0^2}{2\kappa-3}\right)^{-\kappa}\left(1 + \gamma \sqrt{-\epsilon} - \beta\,\epsilon\right)\Bigr] \,,
\nonumber
\end{eqnarray}
where 
\begin{equation}
\label{e7}
A = \frac{\Gamma(\kappa)}{\sqrt{\pi}\sqrt{2\kappa-3}\,\Gamma(\kappa-1/2)} \,,
\end{equation}
together with $\epsilon = v^2/2-\phi,\,\, \sigma = {\rm sgn}(v)$. Moreover,
\begin{equation}
\label{e8}
\frac{\partial^2\phi}{\partial x^2} = n - 1 \,, \quad n = \int_{-\infty}^{\infty} f(v) dv \,.
\end{equation}
Notice that in the Maxwellian limit $\kappa \rightarrow \infty$ one has $A \rightarrow 1/\sqrt{2\pi}$. In addition, in the unperturbed case $\phi = 0, \Psi = 0$, one has $n = 1$. 

\section{Pseudo-potential Method}

\begin{figure}[h]
\begin{center}
\includegraphics[width=4.5in]{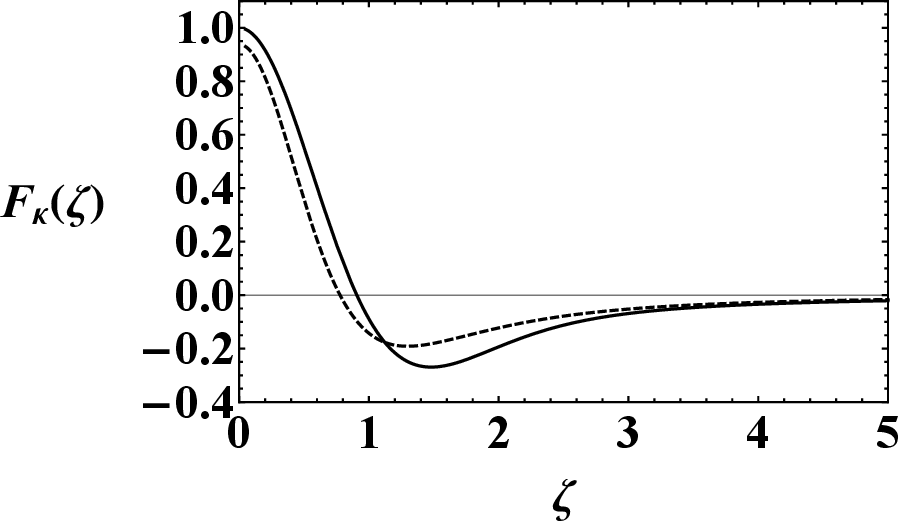}
\caption{This is the behavior of $F_\kappa(\zeta)$ defined in Eq. (\ref{e15}) as a function of $\zeta$ for $\kappa = 2$ (dotted curve), 
and $\kappa = 100$ (continuous curve). }
\label{fig1}
\end{center}
\end{figure}

Our job is to evaluate the electrons number density $n = n(\phi)$ according to 
\begin{eqnarray}
\nonumber 
\frac{n}{A} &=& \Bigr(1 + \frac{k_0^2\Psi}{2}\Bigr)\Bigl[\int_{-\infty}^{-\sqrt{2\phi}} dv\left(1 + \frac{1}{2\kappa-3}(\sqrt{2\epsilon}-v_0)^2\right)^{-\kappa} 
+  \\ &+& \int_{\sqrt{2\phi}}^{\infty} dv\left(1 + \frac{1}{2\kappa-3}(\sqrt{2\epsilon}+v_0)^2\right)^{-\kappa}  
\label{e9}
+ \\ \nonumber &+& \alpha\left(1 + \frac{v_0^2}{2\kappa-3}\right)^{-\kappa}\int_{-\sqrt{2\phi}}^{\sqrt{2\phi}} dv (1 + \gamma\sqrt{-\epsilon} - \beta\epsilon)\Bigr] \,.
\end{eqnarray}
The electrons number density can be obtained from velocity integration followed by Taylor expansion in powers of $\sqrt{\phi}$ as in \cite{Schamel1972, Schamel1986}, or by first Taylor expanding and then performing the velocity integration as in \cite{Schamel1975, Schamel1973, Korn}. In both approaches the result is 
\begin{equation}
\label{e10}
n = 1 + \frac{k_0^2\Psi}{2}+ 2\sqrt{2} A (\alpha-1) \left(1 + \frac{k_0^2\Psi}{2}\right) \left(1 + \frac{v_0^2}{2\kappa-3}\right)^{-\kappa}\sqrt{\phi} + a\,\phi + b\,\phi\sqrt{\phi} + \dots \,,
\end{equation}
valid up to ${\cal O}(\phi^{3/2})$, where 
\begin{eqnarray}
\label{e11}
a &=& \frac{\pi\sqrt{2}}{2}\alpha\gamma A \left(1 + \frac{v_0^2}{2\kappa-3}\right)^{-\kappa} - \frac{1}{2}\left(\frac{\kappa-1}{\kappa-3/2}\right)^2 \frac{d}{d\zeta}Z_{r,\kappa-1}^{*}(\zeta) \,,\\
b &=& \frac{4\sqrt{2} A}{3}\!\left[\alpha\beta \left(1\!+\!\frac{v_0^2}{2\kappa-3}\right)^{-\kappa}\!\!\! + \frac{2\kappa}{(2\kappa-3)^2}[2\kappa(v_0^2-1)+v_0^2+3]\left(1 + \frac{v_0^2}{2\kappa-3}\right)^{-\kappa-2}\right] \,. \nonumber \\
\label{e12}
&\strut&
\end{eqnarray}
In Eq. (\ref{e11}) there is the presence of the real part for real argument of the generalized $\kappa$ plasma dispersion function introduced in \cite{Summers}, 
\begin{equation}
\label{e13}
Z_{\kappa}^{*}(\zeta) = \frac{1}{\pi^{1/2}\kappa^{3/2}}\frac{\Gamma(\kappa+1)}{\Gamma(\kappa-1/2)}\int_{-\infty}^{\infty}\frac{ds}{s-\zeta}\left(1 + \frac{s^2}{\kappa}\right)^{-\kappa-1} \,, \quad {\rm Im}(\zeta) > 0 \,,
\end{equation}
analytically continued for ${\rm Im}(\zeta) < 0$, where the argument is 
\begin{equation}
\label{e14}
\zeta = \left(\frac{\kappa-1}{2\kappa-3}\right)^{1/2}v_0 \,,
\end{equation}
see also \cite{Hellberg2002, Mace, Thorne, Meng, Xue, Summers1994} for properties and applications. As shown in \cite{Mace}, the integral in Eq. (\ref{e13}) can be made single-valued even for non-integer $\kappa$, provided the complex plane is cut and the integration
contour does not cross these cuts. Among other properties, the generalized plasma dispersion function reduces to the usual well known Fried-Conte plasma dispersion function in the Maxwellian limit $\kappa \rightarrow \infty$, which also implies $\zeta \rightarrow v_0/\sqrt{2}$. To express the coefficient $a$ in the form shown in Eq. (\ref{e11}), we employed the property 
\begin{equation}
\label{e15}
-\,\frac{1}{2}\frac{dZ_{\kappa}^{*}(\zeta)}{d\zeta} = 1 - \frac{1}{4\kappa^2} + \left(\frac{\kappa-1/2}{\kappa}\right)\left(\frac{\kappa+1}{\kappa}\right)^{3/2}\zeta\,Z_{\kappa+1}^{*}\left[\left(\frac{\kappa+1}{\kappa}\right)^{1/2}\zeta\right] \equiv F_\kappa(\zeta) \,,
\end{equation}
demonstrated in \cite{Summers}. In what follows, for simplicity of notation, only the real part of $F_\kappa(\zeta)$ defined in Eq. (\ref{e15}) for real argument is considered. Since the derivative of the generalized plasma dispersion function has a role in several of the following steps, we consider 
Figure \ref{fig1} showing aspects of the function $F_\kappa(\zeta)$. Notice the limiting behaviors
\begin{eqnarray}
F_\kappa(0) &=& 1 - \frac{1}{4\kappa^2} > 0 \,, \\
F_\kappa(\zeta) &=& - \,\frac{1}{2}\left(1-\,\frac{1}{2\kappa}\right)\frac{1}{\zeta^2} < 0 \,, \quad \zeta \gg 1 \,. 
\end{eqnarray}
where both inequalities are valid since $\kappa > 3/2$. 


In possession of the electrons number density in terms of $\phi$, it is possible to derive the pseudo-potential $V = V(\phi)$, or Sagdeev potential, from
\begin{equation}
\frac{d^2\phi}{dx^2} = n - 1 = - \,\frac{\partial V}{\partial\phi} \,,
\end{equation}
so that 
\begin{equation}
\label{e16}
- V = \frac{k_0^2\Psi\phi}{2}+ \frac{4\sqrt{2} A}{3} (\alpha-1) \left(1 + \frac{k_0^2\Psi}{2}\right) \left(1 + \frac{v_0^2}{2\kappa-3}\right)^{-\kappa}\phi\sqrt{\phi} + \frac{a\,\phi^2}{2} + \frac{2b\,\phi^2\sqrt{\phi}}{5} + \dots \,,
\end{equation}
correct up to ${\cal O}(\phi^{5/2})$. Notice that the term proportional to $k_0^2\Psi\phi^{3/2}$ usually is not written in the literature, in spite of being of the same order ${\cal O}(\phi^{5/2}) = {\cal O}(\Psi^{5/2})$. This has no consequences, if $\alpha = 1$ (continuous distribution) or if ultimately the analysis is limited to a lower order. With this proviso, the results are fully consistent with \cite{Schamel2015} in the Maxwellian limit $\kappa \rightarrow \infty$. 

A self-consistent solution be it oscillatory or of solitary wave kind requires 
\begin{enumerate}[label=(\roman*)]
\item $V(\phi) < 0$ in the interval $0 < \phi < \Psi$;
\item $V(\Psi) = 0$ \,,
\end{enumerate}
where the later corresponds to zero electric field at the potential maximum. From it we have 
\begin{equation}
\label{e17}
\frac{8\sqrt{2} A}{3\sqrt{\Psi}} (1 - \alpha) \left(1 + \frac{k_0^2\Psi}{2}\right) \left(1 + \frac{v_0^2}{2\kappa-3}\right)^{-\kappa} = k_0^2 +a + \frac{4b}{5}\sqrt{\Psi} \,.
\end{equation}
Equation (\ref{e17}) allows rewriting the pseudo-potential according to 
\begin{eqnarray}
\nonumber 
- V &=& \frac{4\sqrt{2} A}{3\sqrt{\Psi}} (\alpha-1) \left(1 + \frac{k_0^2\Psi}{2}\right) \left(1 + \frac{v_0^2}{2\kappa-3}\right)^{-\kappa}\phi^{3/2}(\sqrt{\Psi}-\sqrt{\phi}) + \\ 
\label{e18}
&+& \frac{k_0^2\phi}{2}(\Psi-\phi) + \frac{2b\phi^2}{5}(\sqrt{\phi}-\sqrt{\Psi}) \,.
\end{eqnarray}
The second term in Re. (\ref{e18}) corresponds to a monochromatic solution $\phi = (\Psi/2)(1 + \cos k_0 x)$ while the last term yields  $\phi = \psi\, {\rm sech}^4(\sqrt{a}x/4)$ solitary wave, taking into account the dispersion relation with $\alpha = 1, k_0 = 0$. These are the same conclusions as from a Maxwellian plasma \cite{Schamel2015}, but with modified coefficients. 

The present main focus is on the strong singularities induced by $\alpha \neq 1$ and $\gamma \neq 0$, the later associated with a non-analytic trapped electrons distribution. Hence we follow the trend of \cite{Schamel2015} and consider small enough amplitudes so that some terms can be neglected in Eqs. (\ref{e17}) and (\ref{e18}), yielding 
\begin{eqnarray}
\label{e19}
&\strut& 
\frac{8\sqrt{2} A}{3\sqrt{\Psi}} (1 - \alpha) \left(1 + \frac{v_0^2}{2\kappa-3}\right)^{-\kappa} = k_0^2 +a  \,, \\
- V &=& \frac{4\sqrt{2} A}{3\sqrt{\Psi}} (\alpha-1)  \left(1 + \frac{v_0^2}{2\kappa-3}\right)^{-\kappa}\phi^{3/2}(\sqrt{\Psi}-\sqrt{\phi}) +  
\frac{k_0^2\phi}{2}(\Psi-\phi)  \,.
\label{e20}
\end{eqnarray}

Equations (\ref{e19}) and (\ref{e20}) are the ultimate tools for our consideration of some special kinds of solutions, all found from the quadrature of the energy-like first integral
\begin{equation}
\label{e21}
\frac{1}{2}\left(\frac{d\phi}{dx}\right)^2 + V(\phi) = 0 \,,
\end{equation}
set to zero without loss of generality taking $V(\Psi) = 0$. Equation (\ref{e19}) is the nonlinear dispersion relation of the problem, relating phase velocity $v_0$, wavenumber $k_0$ and amplitude $\Psi$. The non-analytic contribution from $\gamma \neq 0$ is present in $a$ defined in  Eq.  (\ref{e11}). 

\section{Special solutions}

\subsection{Absence of trapped electrons}

In the case of a void in phase space with no trapped electrons ($\alpha = 0$), further specialized to $k_0 = 0$, one has from Eq. (\ref{e20})
\begin{equation}
V = \frac{4\sqrt{2} A}{3\sqrt{\Psi}} \left(1 + \frac{v_0^2}{2\kappa-3}\right)^{-\kappa}\phi^{3/2}(\sqrt{\Psi}-\sqrt{\phi}) \,.
\end{equation}
Since $V > 0$ in the interval $0 < \phi < \Psi$, it is disqualified as pseudo-potential. 

However, still assuming $\alpha = 0$ but with $k_0 \neq 0$, the nonlinear dispersion relation (\ref{e19}) becomes 
\begin{equation}
\label{e22}
\frac{8\sqrt{2} A}{3\sqrt{\Psi}} \left(1 + \frac{v_0^2}{2\kappa-3}\right)^{-\kappa} = k_0^2 - \frac{1}{2}\left(\frac{\kappa-1}{\kappa-3/2}\right)^2 \frac{d}{d\zeta}Z_{r,\kappa-1}^{*}(\zeta) > 0 \,,
\end{equation}
which is certainly meaningful for sufficiently large $k_0$. Moreover, a small $\Psi$ is assured for large enough $v_0$. The inequality in Eq. (\ref{e22}) also holds, independently of $k_0$, provided $\zeta < \zeta_0$, so that $F_{\kappa-1}(\zeta) > 0$, where $\zeta_0$ is the zero of $F_{\kappa-1}(\zeta)$, or 
$F_{\kappa-1}(\zeta_0) = 0$. In this context a smaller phase velocity would be preferable. 

\begin{figure}[h]
\begin{center}
\includegraphics[width=10.5 cm]{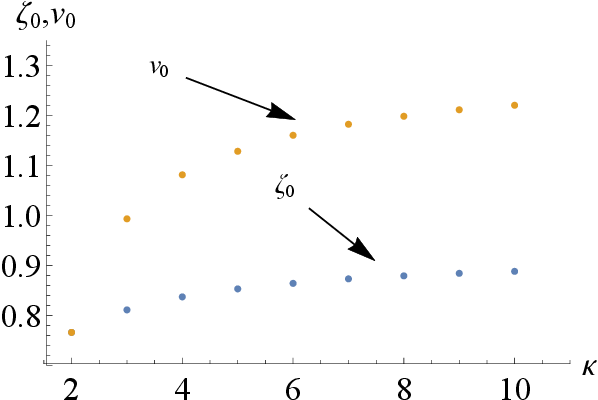}
\caption{Lower: the zero $\zeta_0$ so that $F_{\kappa-1}(\zeta_0) = 0$, as a function of $\kappa$, and the corresponding phase velocity $v_0$ (upper). }
\label{fig2}
\end{center}
\end{figure}

It is relevant to examine the behavior of $\zeta_0$ as a function of $\kappa$. Numerically finding the root of $F_{\kappa-1}(\zeta)$ yields Fig \ref{fig2}, where the corresponding phase velocity is also shown. Asymptotically one has $\zeta_0 \rightarrow v_0/\sqrt{2} = 0.925$ as $\kappa$ increases. Since $\zeta_0$ increases with $\kappa$, the Maxwellian limit allows satisfying the inequality in Eq. (\ref{e22}) irrespective of $k_0$ in an easier way. 

\begin{figure}[h]
\begin{center}
\includegraphics[width=10.5 cm]{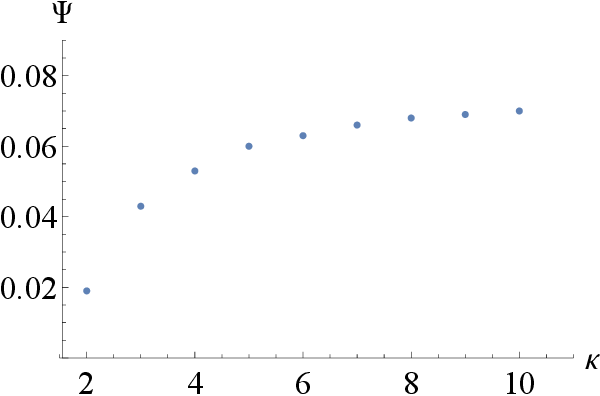}
\caption{Dependence of the amplitude $\Psi$ on the spectral index $\kappa$, from the nonlinear dispersion relation (\ref{e22}), for parameters $k_0 = 2, v_0 = 1$.}
\label{fig3}
\end{center}
\end{figure}

As a rule, the nonlinear dispersion relation (\ref{e22}) is satisfied by larger amplitudes $\Psi$ as $\kappa$ increases, as seen in Fig. \ref{fig3} for $k_0 = 2, v_0 = 1$. Therefore the more superthermal the plasma is, the smaller is the amplitude of the electron hole.  For this set of parameters the Maxwellian limit $\kappa \rightarrow \infty$ is $\Psi = 0.080$.

The Sagdeev potential satisfies 
\begin{eqnarray}
- V &=& - \,\frac{4\sqrt{2} A}{3\sqrt{\Psi}} \left(1 + \frac{v_0^2}{2\kappa-3}\right)^{-\kappa}\phi^{3/2}(\sqrt{\Psi}-\sqrt{\phi}) +  
\frac{k_0^2\phi}{2}(\Psi-\phi)  \nonumber \\
&=&
\frac{k_0^2\sqrt{\Psi}}{2}\,\phi(\sqrt{\Psi}-\sqrt{\phi}) + \frac{1}{4}\left(\frac{\kappa-1}{\kappa-3/2}\right)^2 \frac{d}{d\zeta}Z_{r,\kappa-1}^{*}(\zeta)\,\phi^{3/2}(\sqrt{\Psi}-\sqrt{\phi}) \,,
\label{e23}
\end{eqnarray}
where the nonlinear dispersion relation (\ref{e22}) was used for the last expression.

\begin{figure}[h]
\begin{center}
\includegraphics[width=10.5 cm]{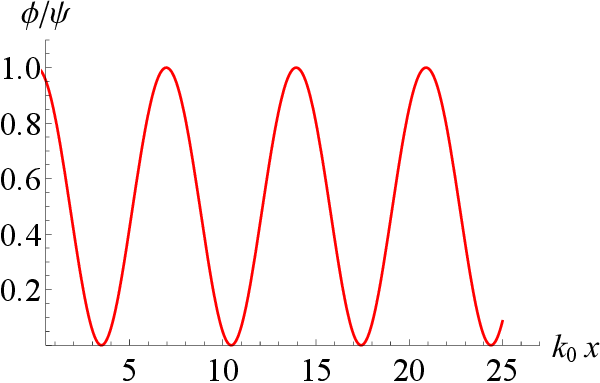}
\caption{Numerical solution of Eq. (\ref{e21}), with pseudo-potential given by Eq. (\ref{e24}), for $\kappa = 2, k_0 = 2, v_0 = 1$.}
\label{fig4}
\end{center}
\end{figure}

The pseudo-potential (\ref{e23}) can be consistent with periodic solutions, which can be seen from the arguments in \cite{Schamel2015}, now adapted to a nonthermal plasma. At the right, most critical border $\phi \rightarrow \Psi^{-}$, one has 
\begin{equation}
\label{e24}
- V = \frac{1}{2}\left(k_0^2 + \frac{1}{2}\left(\frac{\kappa-1}{\kappa-3/2}\right)^2 \frac{d}{d\zeta}Z_{r,\kappa-1}^{*}(\zeta)\right)\Psi^{3/2}(\sqrt{\Psi}-\sqrt{\phi}) \,,
\end{equation}
The right hand side of Eqs. (\ref{e23}) and (\ref{e24}) should be positive to guarantee the existence of a solution, together with a small amplitude, which is assured for large enough $v_0, k_0$, or large phase velocity, say, of order unity, and small wavelength. The overall conclusion is that solitary structures cannot exist under zero trapping conditions, but periodic waves can still be found. Figure \ref{fig4} shows typical nonlinear oscillations from the pseudo-potential (\ref{e23}), with wavelength approximately given by $2\pi/k_0$. The details of these periodic solutions can be shown to be sensitive to $\kappa$, as expected.  

\subsection{Non-analytic trapped electrons distribution}

\begin{figure}[h]
\begin{center}
\includegraphics[width=10.5 cm]{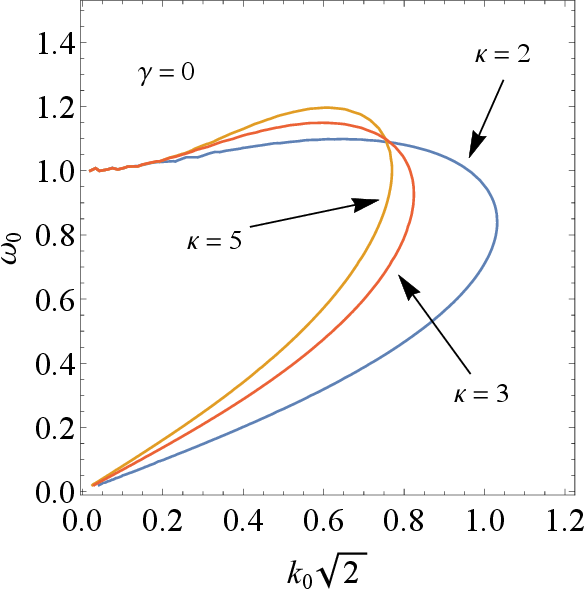}
\caption{The nonlinear dispersion relation (\ref{e25}) where $\omega_0 = k_0 v_0$, with $\gamma = 0$ and different values of $\kappa$, as indicated.}
\label{fig5}
\end{center}
\end{figure}

\begin{figure}[h]
\begin{center}
\includegraphics[width=10.5 cm]{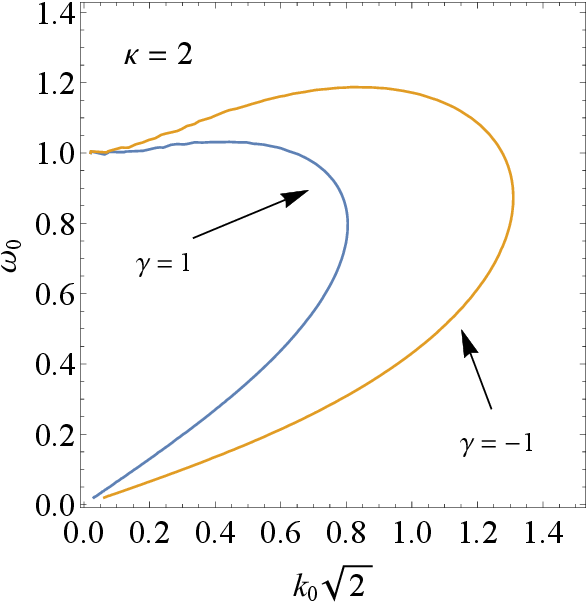}
\caption{The nonlinear dispersion relation (\ref{e25}) where $\omega_0 = k_0 v_0$, with $\kappa = 2$ and different values of $\gamma$, as indicated.}
\label{fig6}
\end{center}
\end{figure}

In the absence of trapping, the parameter $\gamma$ corresponding to a non-analytic trapped electrons distribution obviously plays no role. It is important to examine the influence of $\gamma$ by itself, in the case of a continuous distribution ($\alpha = 1$). In this case the nonlinear dispersion relation is
\begin{equation}
\label{e25}
k_0^2 - \frac{1}{2}\left(\frac{\kappa-1}{\kappa-3/2}\right)^2 \frac{d}{d\zeta}Z_{r,\kappa-1}^{*}(\zeta) = - \frac{\pi\sqrt{2} \gamma A}{2} \left(1 + \frac{v_0^2}{2\kappa-3}\right)^{-\kappa} \,.
\end{equation}
It is similar to (15) of \cite{Schamel2015} and (7) of \cite{Schamel2012}, basically replacing the Fried-Conte function by the generalized plasma dispersion function and $\exp(-v_0^2/2)$ by its finite $\kappa$ power-law version. Though qualitatively the same results from the Maxwellian case are recovered, the value of $\kappa$ influences the details of the dispersion relation. This can be seen in Fig. \ref{fig5}, which is for a non-singular trapped distribution ($\gamma = 0$) and some values of $\kappa$, where $\omega_0 = k_0 v_0$. The well-known thumb curve \cite{Schamel1975} is deformed in accordance with the spectral index, allowing the exploration of smaller wavelengths (bigger $k_0$) the more nonthermal the plasma is. As in the thermal case, one has two branches, the fast
Langmuir branch and the slow electron acoustic branch. The non-analytic case where $\gamma \neq 0$ can gives rise to 
similar deformations, as shown in Fig. \ref{fig6}. We have not found a multitude of dispersion curves as related in \cite{Schamel2015}, where the thermal equivalent of the right-hand side of Eq. (\ref{e25}) was set to constant values, when in fact it is a function of $v_0 = \omega_0/k_0$. 

Notice the sign of $\gamma$ is free. If $\gamma < 0$, a non-negative trapped distribution requires $1 + \gamma\sqrt{\Psi} > 0$. 

\begin{figure}[h]
\begin{center}
\includegraphics[width=10.5 cm]{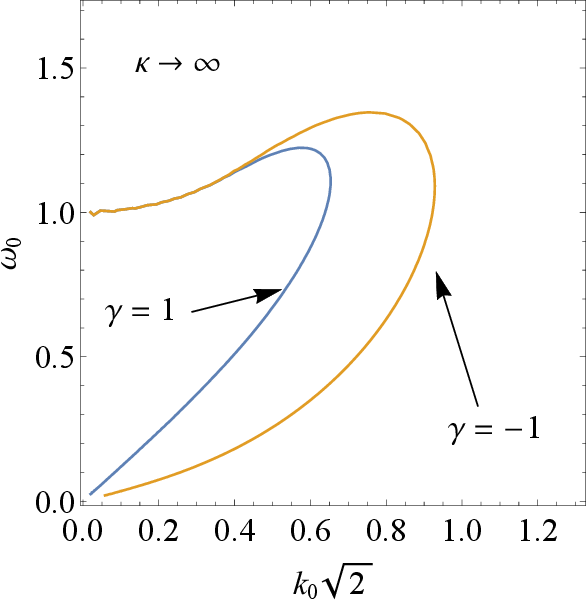}
\caption{The nonlinear dispersion relation (\ref{e25}) where $\omega_0 = k_0 v_0$, with $\kappa \rightarrow \infty $ and different values of $\gamma$, as indicated.}
\label{fig7}
\end{center}
\end{figure}

The Maxwellian limit with non-analytic trapped electrons distribution is shown in Fig. \ref{fig7}. 

\subsection{Surplus of trapped electrons}

In the opposite case of complete absence of trapped electrons, it will be considered $\alpha > 1$, representing a surplus of trapping. For simplicity suppose $k_0 = 0$, so that Eqs. (\ref{e19}) and (\ref{e20}) become 
\begin{eqnarray}
\label{s1}
\frac{\pi\sqrt{2}\alpha\gamma A}{2}\!\!\!\!\!\!  &\strut& \!\!\!\!\!\! \left(1+\frac{v_0^2}{2\kappa-3}\right)^{-\kappa} - \frac{1}{2}\left(\frac{\kappa-1}{\kappa-3/2}\right)^2 \frac{d}{d\zeta}Z_{r,\kappa-1}^{*}(\zeta) = - 2 S \,, \\
\label{s2}
- V &=& S \,\phi^{3/2}(\sqrt{\Psi}-\sqrt{\phi}) \,,
\end{eqnarray}
where 
\begin{equation}
\label{s3}
S =\frac{4\sqrt{2} A (\alpha-1)}{3\sqrt{\Psi}}\left(1+\frac{v_0^2}{2\kappa-3}\right)^{-\kappa} > 0 \,,
\end{equation}
where the last inequality is necessary for $V < 0$ when $0 < \phi < \Psi$. 

In spite of $k_0 = 0$, we haven't a solitary wave but 
\begin{equation}
\phi = \psi\cos^{4}\left(\frac{\sqrt{S}\,x}{2\sqrt{2}}\right) \,.
\label{s4}
\end{equation}
These are the nonthermal equivalent to the results of \cite{Schamel2015} with an excess trapped population, now with a surplus parameter $S$ adapted to the non-Maxwellian background, recovering the previous findings when $\kappa \rightarrow \infty$. 

\begin{figure}[h]
\begin{center}
\includegraphics[width=10.5 cm]{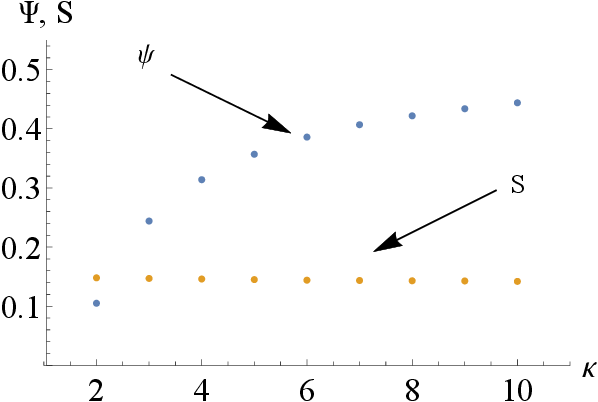}
\caption{Amplitude $\Psi$ and parameter $S$ from the nonlinear dispersion relation (\ref{s1}), for $\gamma = 0, v_0 = 2$, as functions of $\kappa$.}
\label{fig8}
\end{center}
\end{figure}

Assuming for simplicity the analytic case $\gamma = 0$, we can solve Eq. (\ref{s1}) for the amplitude $\Psi$ and $S$ as functions of $\kappa$. This is shown in Fig. \ref{fig8}, calculated for $v_0 = 2$. It is verified that $S$ and hence the oscillations wavelength is not very sensitive to $\kappa$, contrarily to $\Psi$.

\section{Conclusions}

The pseudo-potential approach for a stationary plasma with nonthermal or $\kappa-$dis\-tri\-bu\-ted electrons has been developed, starting from the one-dimensional $\kappa$ distribution and adapting it to be a function of the constants of motion. This is in complete analogy with \cite{Schamel2015} for thermal plasmas. Having the Vlasov equation immediately solved and evaluating the electrons number density in terms of the electrostatic potential up to a certain order, the Poisson equation reduces to a Newtonian-like equation with a conservative potential, or Sagdeev potential. The conditions for solitary wave or oscillatory solutions have been found, leading to a certain nonlinear dispersion relation involving the wave amplitude, the phase velocity and the wavenumber of the propagating structure, besides the spectral index $\kappa$. Following the trend of \cite{Schamel2015}, special attention has been paid to the case of a singular electrons distribution function, allowing for a discontinuity at the separatrix or a non-analytic character. Some special classes of solutions have been discussed. It has been found that in the total absence of trapped electrons only periodic but not solitary wave solutions are allowed. On the other hand, a continuous distribution has a nonlinear dispersion relation modified by the parameter $\gamma$, measuring the strength of the non-analytic character. Finally, in the case of an excess of trapped electrons one can have a periodic solution. This multitude of nonlinear solutions strongly depend on the $\kappa$ parameter. Remarkably, the nonthermal aspects are found from the Maxwellian results replacing the Fried-Conte function by the generalized plasma dispersion function, among other adaptations. 

For simplicity, the allowed non-analytic character was chosen to be of the form proportional to $\sqrt{-\epsilon}$, where $\epsilon < 0$ is the single particle energy of a trapped particle. Certainly, additional choices could be made, such as those with the dependence on $\sqrt{-\epsilon}\ln(-\epsilon)$, as treated several times \cite{Schamel2020a, Krasovsky, Muschietti, Chen}. 

It should be noted that the Vlasov-Poisson model is known to be too restricted for the description of turbulent plasmas with intermittence and eddies \cite{Schamel2020a}, deserving the inclusion of correlations which tend to produce smoother distribution functions. Likewise, the ion dynamics is an essential ingredient ignored in the present communication. Nevertheless, the mathematical tools for more involved electrostatic holes in nonthermal plasmas have been laid down.

\textbf{Acknowledgements} \par 
The author acknowledges financial support by CNPq (Conselho Nacional de De\-sen\-vol\-vi\-men\-to Cient\'{\i}fico e Tecnol\'ogico), Brazil.

\end{document}